\documentclass{amsart}
\usepackage{graphicx}
\newtheorem{thm}{Theorem}[section]

\newcommand{\BP}{{\mathbb P}}

\newcommand{\dsfrac}[2]{{\displaystyle\frac{#1}{#2}}}
\begin{document}

\begin{center}
{\huge \bf {Cubic Pencils and Painlev\'e Hamiltonians}}
\vskip10mm
K.~Kajiwara${}^1$, T.~Masuda${}^2$, M.~Noumi${}^2$, 
Y.~Ohta${}^2$ and Y.~Yamada${}^2$\\
{\normalsize ${}^1$ Graduate School of Mathematics, Kyushu University}\\
{\normalsize ${}^2$ Department of Mathematics, Kobe University}
\end{center}
\vskip20mm
{\bf Abstract}
We present a simple heuristic method to derive the Painlev\'e 
differential equations from the corresponding geometry of rational
surafces. We also give a direct relationship between the cubic pencils and 
Seiberg-Witten curves.
\vskip20mm

\section{Introduction}

For each Painlev\'e equation, there exists an associated rational surface
called the ``space of initial conditions''.
This surface was introduced by Okamoto\cite{Okamoto}, and further studied
by Takano and his collaborators.
By the work of Sioda and Takano\cite{Takano},
the corresponding Painlev\'e equation was characterized 
as the unique Hamiltonian system satisfying certain holomorphy
properties on the surface.  
Hence, in principle, one can recover the Painlev\'e equations 
from geometry.

This geometric approach to the Painlev\'e equations has been
extended to the difference (or discrete) cases, 
from which the difference Painlev\'e equations (and their B\"acklund
transformations) arise naturally as Cremona automorphisms of the 
surfaces\cite{Sakai}. 
Compared with the difference cases, however, the way how the differential
Painlev\'e equations appear is rather indirect.
The known method used so far to recover the differential Painlev\'e equations 
from geometry is either to take suitable continuous limit of discrete ones or
to employ a deformation theory \cite{S-T}.
The aim of this note is to present yet another way, 
which is heuristic but much simpler.

The main idea of our method is to use cubic pencils.
In our previous work\cite{KMNOY}, it is clarified that the cubic pencils 
play the essential role in the discrete Painlev\'e equation. 
It is natural to expect that they are also important
in the differential Painlev\'e equations. Indeed, we find that the cubic
pencils are directly related to the symplectic forms and Hamiltonians.

In Section 2, we explain our method in the case of the sixth Painlev\'e
equation $P_{\rm VI}$. All the other degenerate cases are treated in
Section 3. Finally, a relation of our cubic pencils and the Seiberg-Witten
curves are discussed in Appendix A.

\section{Procedure to obtain Hamiltonian}

In this section, using the sixth Painlev\'e equation $P_{\rm VI}$ as an
example, we explain a procedure to obtain the symplectic 2-form $\omega$
and the Hamiltonian $H$ from the datum of the surface: the configuration
of nine points on $\BP^2$. The parameterization of the points is
borrowed from \cite{Sakai}.

\begin{figure}[h]
\begin{center}\setlength{\unitlength}{0.5mm}
\begin{picture}(180,60)(0,0)
\put(0,0){\line(2,3){40}}
\put(60,0){\line(-2,3){40}}
\put(30,0){\line(0,1){60}}
\put(30,45){\circle*{2}}\put(33,45){189}
\put(10,15){\circle*{2}}\put(13,15){5}
\put(20,30){\circle*{2}}\put(23,30){4}
\put(30,15){\circle*{2}}\put(33,15){3}
\put(30,30){\circle*{2}}\put(33,30){2}
\put(50,15){\circle*{2}}\put(53,15){7}
\put(40,30){\circle*{2}}\put(43,30){6}
\put(70,40){$\longleftarrow$}
\put(100,0){\line(0,1){60}}\put(100,15){\circle*{2}}\put(100,30){\circle*{2}}
\put(120,0){\line(0,1){60}}\put(120,15){\circle*{2}}\put(120,30){\circle*{2}}
\put(140,0){\line(0,1){60}}\put(140,15){\circle*{2}}\put(140,30){\circle*{2}}
\put(160,0){\line(0,1){60}}\put(160,15){\circle*{2}}
\put(90,45){\line(1,0){80}}
\put(102,0){145}\put(102,15){5}\put(102,30){4}
\put(122,0){123}\put(122,15){3}\put(122,30){2}
\put(142,0){167}\put(142,15){7}\put(142,30){6}
\put(162,0){89}\put(162,15){9}
\put(172,45){18}
\end{picture}
\caption{Configuration for $P_{\rm VI}$: In the right diagram, the
 labels $i, ij$ and $ijk$ represent the divisor classes ${\mathcal E}_i$, 
${\mathcal E}_i-{\mathcal E}_j$  and ${\mathcal E}_0-{\mathcal E}_i-{\mathcal E}_j-{\mathcal E}_k$ 
where ${\mathcal E}_0$ is the line in $\BP^2$ and
${\mathcal E}_1, \ldots, {\mathcal E}_9$ are the exceptional divisors.}
\label{fig:P6}
\end{center}
\end{figure}
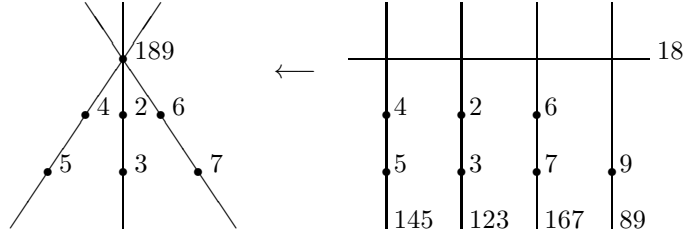

\paragraph{\bf Case $P_{\rm VI}$:} (Fig.\ref{fig:P6}, Add 4)\\

The configuration of the nine points for $P_{\rm VI}$ is
given as follows,
$$
\begin{array}{lll}
P_1=(0:1:0),&
P_2=(1:-a_2:1),&
P_3=(1:-a_1-a_2:1),\\
P_4=(0:0:1),&
P_5=(0:a_3:1),&
P_6=(1:0:0),\\
P_7=(1:a_4:0),&
P_8=((s-1)\epsilon:1:s\epsilon),&
P_9=((s-1)\epsilon:1:s\epsilon-s a_0 \epsilon^2).
\end{array}
$$
Here variables $a_0,a_1,\ldots,a_4$ and $s$ are parameters parameterizing the
configuration. The additional variable 
$\epsilon$ is an infinitesimal parameter introduced in order to
handle some infinitesimally near points.

The configuration for $P_{\rm VI}$ contains a sequence of infinitely near points
$P_{189}=(P_1 \leftarrow P_8 \leftarrow P_9)$. Where $P_i \leftarrow P_j$ means
that the point $P_j$
belongs to the exceptional curve ${\mathcal E}_i \simeq \BP^1$ which is the 
total transform of $P_i$. Here, we represent such
configuration by using an infinitesimal parameter $\epsilon$.
For instance, the condition that a curve
$F(x,y,z)=0$ pass through $P_{18}=(P_1 \leftarrow P_8)$ can be written as
\begin{equation}
F=(s-1)F_x+sF_z=0, \quad ({\rm at} \ P_1)
\end{equation}
or equivalently
\begin{equation}
F(P_8)=F\big((s-1)\epsilon,1,s\epsilon\big)=O(\epsilon^2).
\end{equation}
Similarly, $F(x,y,z)=0$ passes through $P_1, P_8$ and $P_9$ if and only if
\begin{equation}
F(P_9)=F\big((s-1)\epsilon:1:s\epsilon-s a_0 \epsilon^2\big)=O(\epsilon^3).
\end{equation}

Our basic object is a cubic curve passing through the nine points
$P_1,\ldots, P_9$. 
When the parameters $a_i$ are generic, the cubic curve $C_0$ passing
through the nine points is uniquely determined as
\begin{equation}
G=xz(z-x)=0.
\end{equation}
This cubic determines the symplectic form $\omega$:
\begin{equation}
\omega=\frac{x dy\wedge dz+y dz\wedge dx+z dx \wedge dy}{G},
\end{equation}
which can be written as $\omega=df \wedge dg$, with canonical coordinates
\begin{equation}
f=\dsfrac{z}{z-x}, \quad g=\dsfrac{y(z-x)}{xz}.
\end{equation}

When the parameters $a_i$ satisfy the condition 
$\delta=a_0+a_1+2a_2+a_3+a_4=0$, 
the cubic curve passing through the nine points forms a pencil (one
parameter family) Fig. \ref{fig:p6pencil}: 
\begin{figure}[h]
\begin{center}
\includegraphics{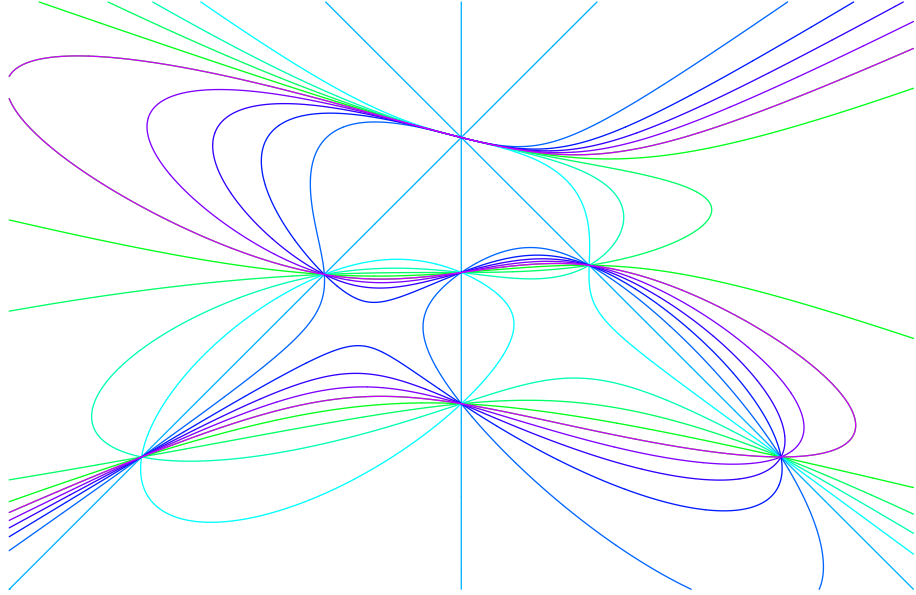}
\caption{Cubic pencil for $P_{\rm VI}$ configuration}
\label{fig:p6pencil}
\end{center}
\end{figure}
\begin{equation}
\lambda F(x,y,z)+\mu G(x,y,z)=0,
\end{equation}
where
\begin{equation}
\begin{array}l
F=-(s-1)y^2z+a_3(s-1)yz^2-a_4s x^2y+a_2(a_1+a_2)x^2z\\[2mm]
\phantom{F=}+sxy^2+(a_1+2a_2+a_3-a_3s+a_4s)xyz. 
\end{array}
\end{equation}
In terms of the canonical variables $f,g$,
the pencil equation $\lambda F+\mu G=0$ can be written as
$\lambda H+\mu=0$ where
\begin{equation}
\begin{array}l
H=f(f-1)(f-s)g^2+\big[(a_1+2a_2)(f-1)f+a_3(s-1)f+a_4s(f-1)\big]g\\[2mm]
\phantom{H=}+a_2(a_1+a_2)(f-1). 
\end{array}
\end{equation}
Note that the choice of $F$ involves the ambiguity such as $F \rightarrow
c_1 F+c_2 G$ where $c_1, c_2$ are constants.
This ambiguity, however, results only in changing $H$ as
$H \rightarrow c_1 H+c_2$.

At this stage, we drop the condition $\delta=0$ by hand. 
We recognize then that $H$ is a Hamiltonian for $P_{\rm VI}$\footnote{
For the autonomous case ($\delta=0$) the pencil is invariant.}, 
namely
\begin{thm}
With the above Hamiltonian $H$, the system of differential equation
\begin{equation}
D_tf=\frac{\partial H}{\partial g}, \quad
D_tg=-\frac{\partial H}{\partial f}, \quad
D_t=s(s-1)\dsfrac{d}{dt}, \quad \dsfrac{ds}{dt}=\delta,
\end{equation}
gives a Hamiltonian form of the sixth Painlev\'e equation $P_{\rm VI}$:
\begin{equation}
\begin{array}l
\dsfrac{d^2f}{dt^2}
=\dsfrac{1}{2}\left(\dsfrac{1}{f}+\dsfrac{1}{f-1}+\dsfrac{1}{f-s}\right)
\left(\dsfrac{df}{dt}\right)^2-\delta
\left(\dsfrac{1}{s}+\dsfrac{1}{s-1}+\dsfrac{1}{f-s}\right)\dsfrac{df}{dt}\\[4mm]
\phantom{\dsfrac{d^2f}{dt^2}=}
+\dsfrac{f(f-1)(f-s)}{s^2(s-1)^2}\left(
\dsfrac{a_1^2}{2}-\dsfrac{a_4^2}{2}\dsfrac{s}{f^2}
+\dsfrac{a_3^2}{2}\dsfrac{s-1}{(f-1)^2}+
\dsfrac{(\delta^2-a_0^2)}{2}\dsfrac{s(s-1)}{(f-s)^2}
\right).
\end{array}
\end{equation}
\end{thm}
 
In the next section, we will show similar results for all the cases in
Table \ref{table:cases}.
\begin{table}
\begin{center}
\begin{tabular}{|c|c|c|c|}
\noalign{\hrule height0.8pt}
Painlev\'e eq.&Sakai's list \cite{Sakai}&configuration&symmetry\\
\noalign{\hrule height0.8pt}
$P_{\rm VI}$&Add 4&$D^{(1)}_4$(Fig.\ref{fig:P6})&$D^{(1)}_4$\\
$P_{\rm V}$&Add 5&$D^{(1)}_5$(Fig.\ref{fig:P5})&$A^{(1)}_3$\\
$P_{\rm III}^{D^{(1)}_6}$&Add 6&$D^{(1)}_6$(Fig.\ref{fig:P3-1})&$(2A_1)^{(1)}$\\
$P_{\rm III}^{D^{(1)}_7}$&Add 7&$D^{(1)}_7$(Fig.\ref{fig:P3-2})&$A^{(1)}_1$\\
$P_{\rm III}^{D^{(1)}_8}$&Add 8&$D^{(1)}_8$(Fig.\ref{fig:P3-3})&${\frak S}_2$\\
$P_{\rm IV}$&Add 9&$E^{(1)}_6$(Fig.\ref{fig:P4})&$A^{(1)}_2$\\
$P_{\rm II}$&Add 10&$E^{(1)}_7$(Fig.\ref{fig:P2})&$A^{(1)}_1$\\
$P_{\rm I}$&Add 11&$E^{(1)}_8$(Fig.\ref{fig:P1})&$-$\\
\noalign{\hrule height0.8pt}
\end{tabular}
\end{center}
\caption{The Painlev\'e equations}
\label{table:cases}
\end{table}

\section{Degenerate cases}

In this section, we consider the degenerate cases Add 5-11 in
\cite{Sakai}.
The constructions are essentially the same as the previous section (Add
4) and we give only the relevant data.

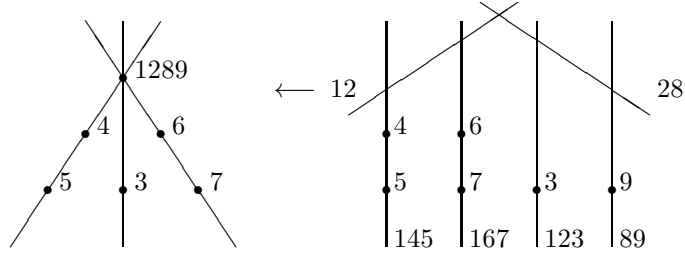
\begin{figure}[h]
\begin{center}\setlength{\unitlength}{0.5mm}
\begin{picture}(180,60)(0,0)
\put(0,0){\line(2,3){40}}
\put(60,0){\line(-2,3){40}}
\put(30,0){\line(0,1){60}}
\put(30,45){\circle*{2}}\put(33,45){1289}
\put(10,15){\circle*{2}}\put(13,15){5}
\put(20,30){\circle*{2}}\put(23,30){4}
\put(30,15){\circle*{2}}\put(33,15){3}
\put(50,15){\circle*{2}}\put(53,15){7}
\put(40,30){\circle*{2}}\put(43,30){6}
\put(70,40){$\longleftarrow$}
\put(100,0){\line(0,1){60}}\put(100,15){\circle*{2}}\put(100,30){\circle*{2}}
\put(120,0){\line(0,1){60}}\put(120,15){\circle*{2}}\put(120,30){\circle*{2}}
\put(140,0){\line(0,1){60}}\put(140,15){\circle*{2}}
\put(160,0){\line(0,1){60}}\put(160,15){\circle*{2}}
\put(90,35){\line(3,2){45}}
\put(125,65){\line(3,-2){45}}
\put(102,0){145}\put(102,15){5}\put(102,30){4}
\put(122,0){167}\put(122,15){7}\put(122,30){6}
\put(142,0){123}\put(142,15){3}
\put(162,0){89}\put(162,15){9}
\put(85,40){12}
\put(172,40){28}
\end{picture}
\caption{Configuration for $P_{\rm V}$}\label{fig:P5}\end{center}
\end{figure}

\paragraph{\bf Case $P_{\rm V}$:} (Fig.\ref{fig:P5}, Add 5)

\noindent
Condition for the cubic: $F(P_i)=0$ ($i=3,4,5,6,7$) and
$F(P_{1289})=0$,
\begin{equation}
\begin{array}l
F(1, -a_2, 1)=F(0, 0, 1)=F(0, a_1, 1)=F(1, 0, 0)=F(1, a_3, 0)=0,\\[2mm]
F(\epsilon, 1, \epsilon + s \epsilon^2 + s(s - a_0)\epsilon^3)=O(\epsilon^4).
\end{array}
\end{equation}
Pencil: $\lambda F+\mu G=0$, ($\delta=a_0+a_1+a_2+a_3=0$)
\begin{equation}
\begin{array}l
F=a_3x^2y-xy^2-a_2sx^2z+(a_1-a_3-s)xyz+y^2z-a_1yz^2,\\[2mm]
G=xz(z-x).
\end{array}
\end{equation}
Hamiltonian $H$ and canonical variables $f,g$:
\begin{equation}
\begin{array}l
H=f(f-1)g(g+s)-(a_1+a_3)fg+a_1g+a_2sf,\\[2mm]
f=\dsfrac{x}{x-z}, \quad g=\dsfrac{y(x-z)}{xz}.
\end{array}
\end{equation}

\begin{thm}
With the above Hamiltonian $H$, the system of differential equation
\begin{equation}
D_tf=\frac{\partial H}{\partial g}, \quad
D_tg=-\frac{\partial H}{\partial f}, \quad
D_t=s\dsfrac{d}{dt}, \quad \dsfrac{ds}{dt}=\delta,
\end{equation}
gives a Hamiltonian form of the fifth Painlev\'e equation $P_{\rm V}$:
$(y=1-1/f)$
\begin{equation}
\begin{array}l
\dsfrac{d^2y}{dt^2}
=\left(\dsfrac{1}{2y}+\dsfrac{1}{y-1}\right)
\left(\dsfrac{dy}{dt}\right)^2-\dsfrac{\delta}{s}\dsfrac{df}{dt}\\[4mm]
\phantom{\dsfrac{d^2y}{dt^2}=}
+\dsfrac{(y-1)^2}{s^2}\left(
\dsfrac{a_1^2}{2}y-\dsfrac{a_3^2}{2}\dsfrac{1}{y}\right)
+(a_0-a_2)\dsfrac{y}{s}-
\dsfrac{1}{2}\dsfrac{y(y+1)}{(y-1)}.
\end{array}
\end{equation}
\end{thm}

\begin{figure}[h]
\begin{center}\setlength{\unitlength}{0.5mm}
\begin{picture}(180,60)(0,0)
\put(0,0){\line(2,3){40}}
\put(60,0){\line(-2,3){40}}
\put(30,0){\line(0,1){60}}
\put(30,45){\circle*{2}}\put(33,45){12389}
\put(10,15){\circle*{2}}\put(13,15){5}
\put(20,30){\circle*{2}}\put(23,30){4}
\put(40,30){\circle*{2}}\put(43,30){6}
\put(50,15){\circle*{2}}\put(53,15){7}
\put(70,40){$\longleftarrow$}
\multiput(100,5)(15,0){5}{\line(0,1){55}}
\put(95,45){\line(1,0){40}}\put(87,45){12}
\put(125,20){\line(1,0){40}}\put(167,20){38}
\put(96,0){145}\put(100,15){\circle*{2}}\put(100,30){\circle*{2}}
\put(102,15){5}\put(102,30){4}
\put(111,0){167}\put(115,15){\circle*{2}}\put(115,30){\circle*{2}}
\put(117,15){7}\put(117,30){6}
\put(128,0){23}
\put(141,0){123}
\put(158,0){89}
\put(160,45){\circle*{2}}\put(162,45){9}
\end{picture}
\caption{Configuration for $P_{\rm III}^{D^{(1)}_6}$}\label{fig:P3-1}\end{center}
\end{figure}
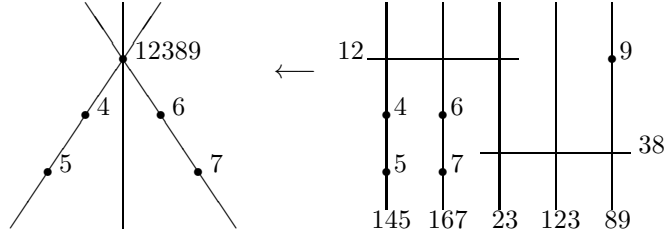

\paragraph{\bf Case $P_{\rm III}^{D^{(1)}_6}$:} 
(Fig.\ref{fig:P3-1}, Add 6)

\noindent
Condition for the cubic: $F(P_i)=0$ ($i=4,5,6,7$) and $F(P_{12389})=0$,
\begin{equation}
\begin{array}l
F(0,0,1)=F(0,a_1,1)=F(1,0,0)=F(1,b_1,0)=0,\\[2mm]
F(\epsilon, 1, \epsilon + s \epsilon^3 + s(b_1 - a_0)\epsilon^4)=O(\epsilon^5).
\end{array}
\end{equation}
Pencil: $\lambda F+\mu G=0$, ($\delta=a_0+a_1=0$)
\begin{equation}
\begin{array}l
F=-b_1x^2y+xy^2+sx^2z+(b_1-a_1)xyz-y^2z+a_1yz^2,\quad G=xz(x-z).
\end{array}
\end{equation}
Hamiltonian $H$ and canonical variables $f,g$:
\begin{equation}
\begin{array}l
H=f^2g^2+\big[f^2-(a_1+b_1)f-s\big]g-a_1f,\\[2mm]
f=\dsfrac{y(z-x)}{xz}, \quad g=\dsfrac{x}{z-x}.
\end{array}
\end{equation}

\begin{thm}
With the above Hamiltonian $H$, the system of differential equation
\begin{equation}
D_tf=\frac{\partial H}{\partial g}, \quad
D_tg=-\frac{\partial H}{\partial f}, \quad
D_t=s\dsfrac{d}{dt}, \quad \dsfrac{ds}{dt}=\delta,
\end{equation}
gives a Hamiltonian form of the third Painlev\'e equation 
$P_{\rm III}^{D^{(1)}_6}$:
\begin{equation}
\dsfrac{d^2f}{dt^2}
=\dsfrac{1}{f}\left(\dsfrac{df}{dt}\right)^2-\dsfrac{\delta}{s}\dsfrac{df}{dt}
+\dsfrac{f^2}{s^2}(f+a_1-b_1)-\dsfrac{1}{f}-\dsfrac{a_0+2 a_1+b_1}{s}.
\end{equation}
\end{thm}

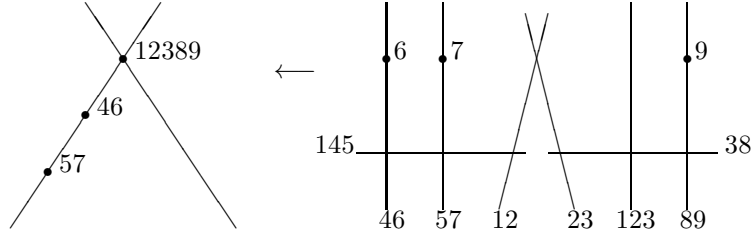
\begin{figure}[h]
\begin{center}\setlength{\unitlength}{0.5mm}
\begin{picture}(180,60)(0,0)
\put(0,0){\line(2,3){40}}
\put(60,0){\line(-2,3){40}}
\put(30,45){\circle*{2}}\put(33,45){12389}
\put(10,15){\circle*{2}}\put(13,15){57}
\put(20,30){\circle*{2}}\put(23,30){46}
\put(70,40){$\longleftarrow$}
\put(100,5){\line(0,1){55}}\put(98,0){46}
\put(115,5){\line(0,1){55}}\put(113,0){57}
\put(130,5){\line(1,4){13}}\put(128,0){12}
\put(150,5){\line(-1,4){13}}\put(148,0){23}
\put(165,5){\line(0,1){55}}\put(161,0){123}
\put(180,5){\line(0,1){55}}\put(178,0){89}
\put(92,20){\line(1,0){45}}\put(81,20){145}
\put(143,20){\line(1,0){45}}\put(190,20){38}
\put(100,45){\circle*{2}}\put(102,45){6}
\put(115,45){\circle*{2}}\put(117,45){7}
\put(180,45){\circle*{2}}\put(182,45){9}
\end{picture}
\caption{Configuration for $P_{\rm III}^{D^{(1)}_7}$}\label{fig:P3-2}\end{center}
\end{figure}

\paragraph{\bf Case $P_{\rm III}^{D^{(1)}_7}$:}(Fig.\ref{fig:P3-2}, Add 7)

\noindent
Condition for the cubic: $F(P_{46})=F(P_{57})=F(P_{12389})=0$,
\begin{equation}
\begin{array}l
F(\epsilon,0,1)=O(\epsilon^2),\quad
F(\epsilon,1+a_1 \epsilon,1)=O(\epsilon^2), \quad
F(\epsilon, 1, s \epsilon^3 - a_0 s \epsilon^4)=O(\epsilon^5).
\end{array}
\end{equation}
Pencil: $\lambda F+\mu G=0$, ($\delta=a_0+a_1=0$)
\begin{equation}
F=-s x^3-a_1 x y z+y^2 z-y z^2,\quad G=x^2 z.
\end{equation}
Hamiltonian $H$ and canonical variables $f,g$:
\begin{equation}
\begin{array}l
H=f^2g^2+(a_1f+s)g-f,\\[2mm]
f=\dsfrac{yz}{x^2}, \quad g=-\dsfrac{x}{z}.
\end{array}
\end{equation}

\begin{thm}
With the above Hamiltonian $H$, the system of differential equation
\begin{equation}
D_tf=\frac{\partial H}{\partial g}, \quad
D_tg=-\frac{\partial H}{\partial f}, \quad
D_t=s\dsfrac{d}{dt}, \quad \dsfrac{ds}{dt}=\delta,
\end{equation}
gives a Hamiltonian form of the third Painlev\'e equation 
$P_{\rm III}^{D^{(1)}_7}$:
\begin{equation}
\dsfrac{d^2f}{dt^2}
=\dsfrac{1}{f}\left(\dsfrac{df}{dt}\right)^2-\dsfrac{\delta}{s}\dsfrac{df}{dt}
+2\dsfrac{f^2}{s^2}-\dsfrac{1}{f}+\dsfrac{a_0}{s}.
\end{equation}
\end{thm}

\begin{figure}[h]
\begin{center}\setlength{\unitlength}{0.5mm}
\begin{picture}(180,60)(0,0)
\put(0,0){\line(2,3){40}}
\put(60,0){\line(-2,3){40}}
\put(30,45){\circle*{2}}\put(33,45){12389}
\put(10,15){\circle*{2}}\put(13,15){4567}
\put(70,40){$\longleftarrow$}
\put(100,5){\line(0,1){55}}\put(98,0){45}
\put(115,5){\line(0,1){55}}\put(113,0){67}
\put(130,5){\line(0,1){55}}\put(126,0){145}
\put(150,5){\line(0,1){55}}\put(148,0){23}
\put(165,5){\line(0,1){55}}\put(161,0){123}
\put(180,5){\line(0,1){55}}\put(178,0){89}
\put(92,20){\line(1,0){45}}\put(81,20){56}
\put(143,20){\line(1,0){45}}\put(190,20){38}
\put(125,45){\line(1,0){30}}\put(155,45){12}
\put(115,45){\circle*{2}}\put(117,45){7}
\put(180,45){\circle*{2}}\put(182,45){9}
\end{picture}
\caption{Configuration for $P_{\rm III}^{D^{(1)}_8}$}\label{fig:P3-3}\end{center}
\end{figure}
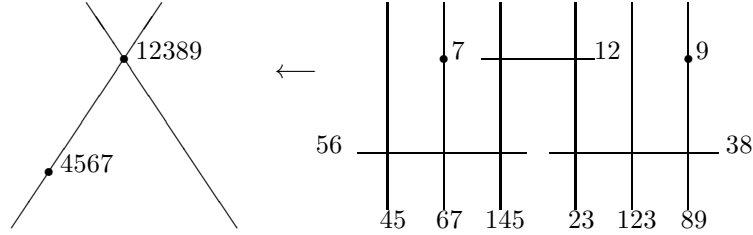

\paragraph{\bf Case $P_{\rm III}^{D^{(1)}_8}$:}
(Fig.\ref{fig:P3-3}, Add 8)

\noindent
Condition for the cubic: $F(P_{4567})=F(P_{12389})=0$,
\begin{equation}
\begin{array}l
F(\epsilon^2,\epsilon,1)=O(\epsilon^4),\quad
F(\epsilon, 1, s \epsilon^3 - a s \epsilon^4)=O(\epsilon^5).
\end{array}
\end{equation}
Pencil: $\lambda F+\mu G=0$, ($\delta=a=0$)
\begin{equation}
\begin{array}l
F=-sx^3+y^2z-xz^2,\quad G=x^2 z.
\end{array}
\end{equation}
Hamiltonian $H$ and canonical variables $f,g$:
\begin{equation}
\begin{array}l
H=f^2g^2-f-\dsfrac{s}{f},\\[3mm]
f=\dsfrac{z}{x}, \quad g=-\dsfrac{y}{z}.
\end{array}
\end{equation}

\begin{thm}
With the above Hamiltonian $H$, the system of differential equation
\begin{equation}
D_tf=\frac{\partial H}{\partial g}, \quad
D_tg=-\frac{\partial H}{\partial f}, \quad
D_t=s\dsfrac{d}{dt}, \quad \dsfrac{ds}{dt}=\delta,
\end{equation}
gives a Hamiltonian form of the third Painlev\'e equation 
$P_{\rm III}^{D^{(1)}_8}$:
\begin{equation}
\dsfrac{d^2f}{dt^2}
=\dsfrac{1}{f}\left(\dsfrac{df}{dt}\right)^2-\dsfrac{\delta}{s}\dsfrac{df}{dt}
+2\dsfrac{f^2}{s^2}-\dsfrac{2}{s}.
\end{equation}
\end{thm}

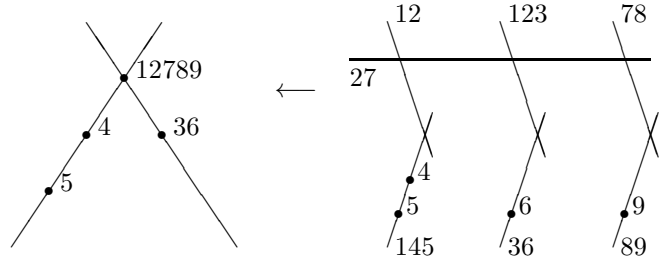
\begin{figure}[h]
\begin{center}\setlength{\unitlength}{0.5mm}
\begin{picture}(180,60)(0,0)
\put(0,0){\line(2,3){40}}
\put(60,0){\line(-2,3){40}}
\put(30,45){\circle*{2}}\put(33,45){12789}
\put(10,15){\circle*{2}}\put(13,15){5}
\put(20,30){\circle*{2}}\put(23,30){4}
\put(40,30){\circle*{2}}\put(43,30){36}
\put(70,40){$\longleftarrow$}
\put(100,0){\line(1,3){12}}\put(100,60){\line(1,-3){12}}
\put(130,0){\line(1,3){12}}\put(130,60){\line(1,-3){12}}
\put(160,0){\line(1,3){12}}\put(160,60){\line(1,-3){12}}
\put(90,50){\line(1,0){80}}\put(90,43){27}
\put(102,60){12}\put(132,60){123}\put(162,60){78}
\put(102,-2){145}\put(132,-2){36}\put(162,-2){89}
\put(106,18){\circle*{2}}\put(108,18){4}
\put(103,9){\circle*{2}}\put(105,9){5}
\put(133,9){\circle*{2}}\put(135,9){6}
\put(163,9){\circle*{2}}\put(165,9){9}
\end{picture}
\caption{Configuration for $P_{\rm IV}$}\label{fig:P4}\end{center}
\end{figure}

\paragraph{\bf Case $P_{\rm IV}$:}(Fig.\ref{fig:P4}, Add 9)

\noindent
Condition for the cubic: $F(P_4)=F(P_5)=F(P_{36})=F(P_{12789})=0$,
\begin{equation}
\begin{array}l
F(0,0,1)=F(0,a_1,1)=0,\\[2mm]
F(1,-a_2\epsilon,\epsilon)=O(\epsilon^2), \quad 
F(\epsilon, 1, \epsilon^2 + s \epsilon^3 + (s^2 - a_0)\epsilon^4)=O(\epsilon^5).
\end{array}
\end{equation}
Pencil: $\lambda F+\mu G=0$, ($\delta=a_0+a_1+a_2=0$)
\begin{equation}
\begin{array}l
F=-x^2y - a_2x^2z - sxyz + y^2z - a_1yz^2,\quad G=xz^2.
\end{array}
\end{equation}
Hamiltonian $H$ and canonical variables $f,g$:
\begin{equation}
\begin{array}l
H=fg(g-f-s)-a_2f-a_1g,\\[2mm]
f=\dsfrac{x}{z}, \quad g=\dsfrac{y}{x}.
\end{array}
\end{equation}

\begin{thm}
With the above Hamiltonian $H$, the system of differential equation
\begin{equation}
D_tf=\frac{\partial H}{\partial g}, \quad
D_tg=-\frac{\partial H}{\partial f}, \quad
D_t=\dsfrac{d}{dt}, \quad \dsfrac{ds}{dt}=\delta,
\end{equation}
gives a Hamiltonian form of the fourth Painlev\'e equation $P_{\rm IV}$:
\begin{equation}
\dsfrac{d^2f}{dt^2}
=\dsfrac{1}{2f}\left(\dsfrac{df}{dt}\right)^2+\dsfrac{3}{2}f^3+2sf^2
+\dsfrac{1}{2}\big[s^2+2(a_2-a_0)\big]f-\dsfrac{a_1^2}{2f}.
\end{equation}
\end{thm}

\begin{figure}[h]
\begin{center}\setlength{\unitlength}{0.5mm}
\begin{picture}(180,60)(0,0)
\put(0,0){\line(2,3){40}}
\put(60,0){\line(-2,3){40}}
\put(30,45){\circle*{2}}\put(33,45){1236789}
\put(10,15){\circle*{2}}\put(13,15){5}
\put(20,30){\circle*{2}}\put(23,30){4}
\put(70,40){$\longleftarrow$}
\put(100,5){\line(0,1){55}}\put(98,0){89}
\put(115,5){\line(0,1){55}}\put(113,0){67}
\put(130,5){\line(0,1){55}}\put(126,0){123}
\put(145,5){\line(0,1){55}}\put(143,0){23}
\put(160,5){\line(0,1){55}}\put(156,0){145}
\put(93,20){\line(1,0){30}}\put(87,20){78}
\put(110,45){\line(1,0){40}}\put(106,45){36}
\put(137,20){\line(1,0){30}}\put(170,20){12}
\put(160,40){\circle*{2}}\put(162,40){4}
\put(160,30){\circle*{2}}\put(162,30){5}
\put(100,30){\circle*{2}}\put(102,30){9}
\end{picture}
\caption{Configuration for $P_{\rm II}$}\label{fig:P2}\end{center}
\end{figure}
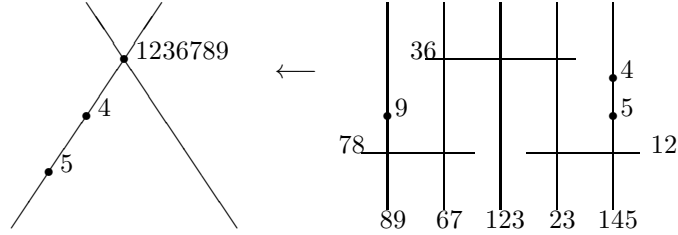

\paragraph{\bf Case $P_{\rm II}$:}(Fig.\ref{fig:P2}, Add 10)

\noindent
Condition for the cubic: $F(P_4)=F(P_5)=F(P_{1236789})=0$,
\begin{equation}
\begin{array}l
F(0,0,1)=F(0,a_1,1)=0,\quad
F(\epsilon, 1, \epsilon^3 - s \epsilon^5-a_0 \epsilon^6)=O(\epsilon^7).
\end{array}
\end{equation}
Pencil: $\lambda F+\mu G=0$, ($\delta=a_0+a_1=0$)
\begin{equation}
F=x^3-sx^2z-y^2z+a_1yz^2,\quad G=x z^2.
\end{equation}
Hamiltonian $H$ and canonical variables $f,g$:
\begin{equation}
H=g^2+(f^2+s)g+a_1f, \quad f=\dsfrac{y}{x}, \quad g=-\dsfrac{x}{z}.
\end{equation}

\begin{thm}
With the above Hamiltonian $H$, the system of differential equation
\begin{equation}
D_tf=\frac{\partial H}{\partial g}, \quad
D_tg=-\frac{\partial H}{\partial f}, \quad
D_t=\dsfrac{d}{dt}, \quad \dsfrac{ds}{dt}=\delta,
\end{equation}
gives a Hamiltonian form of the second Painlev\'e equation $P_{\rm II}$:
\begin{equation}
\dsfrac{d^2f}{dt^2}=2f^3+2sf+(a_0-a_1).
\end{equation}
\end{thm}

\paragraph{\bf Case $P_{\rm I}$}(Fig.\ref{fig:P1}, Add 11)
\begin{figure}[h]
\begin{center}\setlength{\unitlength}{0.5mm}
\begin{picture}(180,60)(0,0)
\put(0,20){\line(1,0){40}}
\put(20,20){\circle*{2}}\put(20,22){123456789}
\put(70,40){$\longleftarrow$}
\put(100,5){\line(0,1){55}}\put(98,0){89}
\put(115,5){\line(0,1){55}}\put(113,0){67}
\put(130,5){\line(0,1){55}}\put(128,0){45}
\put(145,5){\line(0,1){55}}\put(141,0){123}
\put(160,5){\line(0,1){55}}\put(158,0){23}
\put(93,20){\line(1,0){30}}\put(87,20){78}
\put(108,45){\line(1,0){30}}\put(106,45){56}
\put(123,30){\line(1,0){45}}\put(170,30){34}
\put(153,50){\line(1,0){15}}\put(170,50){12}
\put(100,30){\circle*{2}}\put(102,30){9}
\end{picture}
\caption{Configuration for $P_{\rm I}$}\label{fig:P1}\end{center}
\end{figure}

\noindent
Condition for the cubic: $F(P_{123456789})=0$,
\begin{equation}
F(\epsilon, 1, \epsilon^3 +s \epsilon^7+a \epsilon^8)=O(\epsilon^9).
\end{equation}
Pencil: $\lambda F+\mu G=0$: ($\delta=a=0$)
\begin{equation}
F=-x^3+y^2z-sxz^2,\quad G=z^3.
\end{equation}
Hamiltonian $H$ and canonical variables $f,g$:
\begin{equation}
H=g^2-f^3-sf, \quad f=\dsfrac{x}{z}, \quad g=\dsfrac{y}{z}.
\end{equation}

\begin{thm}
With the above Hamiltonian $H$, the system of differential equation
\begin{equation}
D_tf=\frac{\partial H}{\partial g}, \quad
D_tg=-\frac{\partial H}{\partial f}, \quad
D_t=\dsfrac{d}{dt}, \quad \dsfrac{ds}{dt}=\delta,
\end{equation}
gives a Hamiltonian form of the first Painlev\'e equation $P_{\rm I}$:
\begin{equation}
\dsfrac{d^2f}{dt^2}=6 f^2+2 s.
\end{equation}
\end{thm}

\appendix
\section{Relation to Seiberg-Witten curves}

It may be interesting to note that the cubic pencils we considered in this paper
are directly related with the Seiberg-Witten curves
appearing in the ${\mathcal N}=2$ supersymmetric
gauge theory with ${\rm SU}(2)$ gauge group.
The following is the Seiberg-Witten curves given in \cite{SW} and
\cite{AD} (with some parameters rescaled).
\begin{equation}\label{eq:SWcurves}
\begin{array}{ll}
D_8:&
y^2=x^3-ux^2+2 \Lambda_0^4 x.\\[1mm]
D_7:&
y^2=x^2(x-u)+2m_1\Lambda_1^3x-\Lambda_1^6.\\[1mm]
D_6:&
y^2=(x^2-\Lambda_2^4)(x-u)+2m_1m_2\Lambda_2^2x-(m_1^2+m_2^2)\Lambda_2^4.\\[1mm]
D_5:&
y^2=x^2(x-u)-\Lambda_3^2(x-u)^2-
\displaystyle\sum_{i=1}^3m_i^2\Lambda_3^2(x-u)\\[1mm]
&\phantom{y^2=}+2m_1m_2m_3\Lambda_3x-
\displaystyle\sum_{1\leq i<j\leq 3}m_i^2m_j^2\Lambda_3^2.\\[1mm]
D_4:&
y^2=x(x-\alpha u)(x-\beta u)-\dsfrac{1}{4}(\alpha-\beta)^2u_2x^2\\[1mm]
&\phantom{y^2=}-\big(\dsfrac{1}{4}(\alpha-\beta)^2\alpha\beta u_4
-\dsfrac{1}{2}\alpha\beta(\alpha^2-\beta^2)s_4\big)x\\[1mm]
&\phantom{y^2=}-(\alpha-\beta)\alpha^2\beta^2s_4 u
-\dsfrac{1}{4}(\alpha-\beta)^2\alpha^2\beta^2u_6,\\[1mm]
&\displaystyle u_{2}=\sum_{i=1}^4 m_i^2, \quad
u_{4}=\sum_{1\leq i<j\leq 4}m_{i}^2m_{j}^2,\quad
u_{6}=\sum_{1\leq i<j<k\leq 4}m_{i}^2m_{j}^2m_{k}^2,\\[1mm]
&\displaystyle s_{4}=\prod_{i=1}^{4}m_i, \quad
\alpha=-\vartheta_3(\tau)^4, \quad \beta=-\vartheta_2(\tau)^4.\\[1mm]
E_8:&y^2=x^3-2M x-u.\\[1mm]
E_7:&y^2=x^3-2ux-2Mu+M^3-4m_1^2.\\[1mm]
E_6:&y^2=x^3-2(Mu+c_2)x-u^2-\dsfrac{M^3}{3}u+\dsfrac{M^6}{108}
-\dsfrac{2M^2}{3}c_2+\dsfrac{8}{3}c_3,\\[1mm]
&c_k=m_1^k+m_2^k+m_3^k \quad (c_1=0).
\end{array}
\end{equation}

The correspondence between our cubic pencils and the above Seiberg-Witten
curves is a direct consequence of their definition/construction\cite{GMS}
\cite{MY}\cite{ES}.
In fact, by compairing the Weierstrass canonical form of both curves,
the relations of the parameters are explicitly determined as in
Table \ref{table:PL-SW}.

\begin{table}[h]
\begin{center}
\begin{tabular}{|c|c|c|}
\noalign{\hrule height0.8pt}
Painlev\'e & SW curve & Relation of parameters\\
\noalign{\hrule height0.8pt}
$P_{\rm I}$ & $E_8$ & $s=-2M$\\
\noalign{\hrule}
$P_{\rm II}$ & $E_7$ & $ a_1=4 m_1, \quad s=-3M$\\
\noalign{\hrule}
$P_{\rm IV}$ & $E_6$ & $a_1=2(m_1-m_2), \quad a_2=2(m_2-m_3), \quad s=2M$\\
\noalign{\hrule}
${P_{\rm III}}^{D^{(1)}_8}$ & $D_8$ & $s=2 \Lambda_0^4$\\
\noalign{\hrule}
${P_{\rm III}}^{D^{(1)}_7}$ & $D_7$ & $a_1=2 m_1, \quad s=2\Lambda_1^3$\\
\noalign{\hrule}
${P_{\rm III}}^{D^{(1)}_6}$ & $D_6$ & $a_1=m_1-m_2, 
\quad b_1=m_1+m_2, \quad s=-2\Lambda_2^2$\\
\noalign{\hrule}
$P_{\rm V}$ & $D_5$ & $a_1=-(m_1+m_3), \quad a_2=m_1+m_2, \quad a_3=m_3-m_1,
\quad s=2 \Lambda_3$\\
\noalign{\hrule}
$P_{\rm VI}$ & $D_4$ & $
a_1=m_3+m_4, \quad a_2=m_2-m_3, \quad a_3=m_1-m_2, \quad
a_4=m_3-m_4, \quad s=\dsfrac{\beta}{\alpha} $\\
\noalign{\hrule height0.8pt}
\end{tabular}
\end{center}
\caption{Painlev\'e equation and Seiberg-Witten curve}
\label{table:PL-SW}
\end{table}



\begin{thebibliography}{99}
\bibitem{Okamoto}
K.Okamoto,
\textit{Sur les feuilletages associ\'es aux \'equation du second ordre
\`a points critiques fixes de P. Painlev\'e},
Japan. J. Math. {\bf 5} (1979) 1-79.
%
\bibitem{Takano}
T.Sioda and K.Takano,
\textit{On Some Hamiltonian Structures of Painlev\'e Systems,I},
Funkcial. Ekvac, {\bf 40} (1997) 271-291.
%
\bibitem{Sakai}
H.Sakai, 
\textit{Rational surfaces with affine root systems and geometry of
the Painlev\'e equations},
Commun. Math. Phys. {\bf 220} (2001) 165-221.
%
\bibitem{S-T}
M.Saito, T.Takebe and H.Terajima,
\textit{Deformation of Okamoto-Painlev\'e pairs and Painlev\'e equations},
J. Algebraic Geom. {\bf 11} (2002) 311--362.\\
H.Terajima,
\textit{Local cohomology of generalized Okamoto-Painlev\'e pairs and
		Painlev\'e equations},
Japan. J. Math. (N.S.) {\bf 28} (2002) 137--162.
%
\bibitem{KMNOY}
K.Kajiwara, T.Masuda, M.Noumi, Y.Ohta and Y.Yamada,
\textit{${}_{10}E_{9}$ solutions to the elliptic Painlev\'e equation}, 
J.\,Phys.\,A:Math.Gen. {\bf 36} (2003) L263--L272.
%
\bibitem{SW}
N.Seiberg and E.Witten,
\textit{Monopoles, Duality and Chiral Symmetry Breaking in $N=2$
		Supersymmetric QCD},
Nuclear Phys. {\bf B431} (1994) 484--550.
%
\bibitem{AD}
P.C.Argyres, M.R.Plesser, N.Seiberg and E.Witten, 
\textit{New $N=2$ Superconformal Field Theories in Four Dimensions},
Nuclear Phys. {\bf B461} (1996) 71--84. 
%
\bibitem{GMS}
O.J.Ganor, D.R.Morrison and N.Seiberg, 
\textit{Branes, Calabi-Yau Spaces, and Troidal Compactification of the
		$N=1$ Six-Dimensional $E_8$ Theory},
Nuclear Phys. {\bf B487} (1997) 93--127. 
%
\bibitem{MY}
S.Mizoguchi and Y.Yamada,
\textit{$W(E_{10})$ Symmetry, $M$-theory and Painlev\'e equations},
Phys. Lett. {\bf B537} (2002) 130-140.
%
\bibitem{ES}
T.Eguchi and K.Sakai,
\textit{Seiberg-Witten Curve for $E$-String Theory Revisited},
Adv. Theor. Math. Phys. {\bf 7} (2003) 421-457.
%
\end{thebibliography}
\end{document}